\documentclass[
aps,prd,twocolumn,superscriptaddress,nofootinbib,
amsmath,amssymb]{revtex4-2}

\usepackage{orcidlink}
\usepackage{bm}
\usepackage{dcolumn}
\usepackage{graphicx}

\usepackage{adsmacros}

\newcommand{\Tr}{\mathrm{Tr}}

\newcommand{\lya}{Ly$\alpha$}

\newcommand{\partderiv}[2]{\frac{\partial #1}{\partial #2}}

\newcommand{\poned}{$P_{\mathrm{1D}}$}
\newcommand{\bispeconed}{$B_{\kappa, \mathrm{Ly}\alpha}$}
\newcommand{\skm}{s~km$^{-1}$}

\newcommand{\mpch}{Mpc~$h^{-1}$}
\newcommand{\hmpc}{$h$~Mpc$^{-1}$}

\begin{document}

\preprint{PRD}

\title{CMB lensing and \lya\ forest cross bispectrum from DESI's first-year quasar sample}


\author{Naim~G{\" o}ksel~\surname{Kara{\c c}ayl{\i}}\orcidlink{0000-0001-7336-8912}}
\email{karacayli.1@osu.edu}
\affiliation{Center for Cosmology and AstroParticle Physics, The Ohio State University, 191 West Woodruff Avenue, Columbus, OH 43210, USA}
\affiliation{Department of Astronomy, The Ohio State University, 4055 McPherson Laboratory, 140 W 18th Avenue, Columbus, OH 43210, USA}
\affiliation{Department of Physics, The Ohio State University, 191 West Woodruff Avenue, Columbus, OH 43210, USA}
\author{Paul~Martini\orcidlink{0000-0002-4279-4182}}
\affiliation{Center for Cosmology and AstroParticle Physics, The Ohio State University, 191 West Woodruff Avenue, Columbus, OH 43210, USA}
\affiliation{Department of Astronomy, The Ohio State University, 4055 McPherson Laboratory, 140 W 18th Avenue, Columbus, OH 43210, USA}
\author{David~H.~Weinberg}
\affiliation{Department of Astronomy, The Ohio State University, 4055 McPherson Laboratory, 140 W 18th Avenue, Columbus, OH 43210, USA}
\author{Simone~Ferraro}
\affiliation{Lawrence Berkeley National Laboratory, 1 Cyclotron Road, Berkeley, CA 94720, USA}
\affiliation{Department of Astronomy, University of California, Berkeley, 110 Sproul Hall \#5800 Berkeley, CA 94720, USA}
\author{Roger~\surname{de~Belsunce}}
\affiliation{Lawrence Berkeley National Laboratory, 1 Cyclotron Road, Berkeley, CA 94720, USA}
\author{J.~Aguilar}
\affiliation{Lawrence Berkeley National Laboratory, 1 Cyclotron Road, Berkeley, CA 94720, USA}
\author{S.~Ahlen}
\affiliation{Physics Dept., Boston University, 590 Commonwealth Avenue, Boston, MA 02215, USA}
\author{E.~Armengaud}
\affiliation{IRFU, CEA, Universit\'{e} Paris-Saclay, F-91191 Gif-sur-Yvette, France}
\author{D.~Brooks}
\affiliation{Department of Physics \& Astronomy, University College London, Gower Street, London, WC1E 6BT, UK}
\author{T.~Claybaugh}
\affiliation{Lawrence Berkeley National Laboratory, 1 Cyclotron Road, Berkeley, CA 94720, USA}
\author{A.~\surname{de la Macorra}}
\affiliation{Instituto de F\'{\i}sica, Universidad Nacional Aut\'{o}noma de M\'{e}xico,  Cd. de M\'{e}xico  C.P. 04510,  M\'{e}xico}
\author{B.~Dey}
\affiliation{Department of Physics \& Astronomy and Pittsburgh Particle Physics, Astrophysics, and Cosmology Center (PITT PACC), University of Pittsburgh, 3941 O'Hara Street, Pittsburgh, PA 15260, USA}
\author{P.~Doel}
\affiliation{Department of Physics \& Astronomy, University College London, Gower Street, London, WC1E 6BT, UK}
\author{K.~Fanning}
\affiliation{Kavli Institute for Particle Astrophysics and Cosmology, Stanford University, Menlo Park, CA 94305, USA}
\affiliation{SLAC National Accelerator Laboratory, Menlo Park, CA 94305, USA}
\author{J.~E.~Forero-Romero}
\affiliation{Departamento de F\'isica, Universidad de los Andes, Cra. 1 No. 18A-10, Edificio Ip, CP 111711, Bogot\'a, Colombia}
\affiliation{Observatorio Astron\'omico, Universidad de los Andes, Cra. 1 No. 18A-10, Edificio H, CP 111711 Bogot\'a, Colombia}
\author{S.~\surname{Gontcho A Gontcho}}
\affiliation{Lawrence Berkeley National Laboratory, 1 Cyclotron Road, Berkeley, CA 94720, USA}
\author{A.~X.~Gonzalez-Morales}
\affiliation{Consejo Nacional de Ciencia y Tecnolog\'{\i}a, Av. Insurgentes Sur 1582. Colonia Cr\'{e}dito Constructor, Del. Benito Ju\'{a}rez C.P. 03940, M\'{e}xico D.F. M\'{e}xico}
\affiliation{Departamento de F\'{i}sica, Universidad de Guanajuato - DCI, C.P. 37150, Leon, Guanajuato, M\'{e}xico}
\author{G.~Gutierrez}
\affiliation{Fermi National Accelerator Laboratory, PO Box 500, Batavia, IL 60510, USA}
\author{J.~Guy}
\affiliation{Lawrence Berkeley National Laboratory, 1 Cyclotron Road, Berkeley, CA 94720, USA}
\author{K.~Honscheid}
\affiliation{Center for Cosmology and AstroParticle Physics, The Ohio State University, 191 West Woodruff Avenue, Columbus, OH 43210, USA}
\affiliation{Department of Physics, The Ohio State University, 191 West Woodruff Avenue, Columbus, OH 43210, USA}
\author{D.~Kirkby}
\affiliation{Department of Physics and Astronomy, University of California, Irvine, 92697, USA}
\author{T.~Kisner}
\affiliation{Lawrence Berkeley National Laboratory, 1 Cyclotron Road, Berkeley, CA 94720, USA}
\author{A.~Kremin}
\affiliation{Lawrence Berkeley National Laboratory, 1 Cyclotron Road, Berkeley, CA 94720, USA}
\author{A.~Lambert}
\affiliation{Lawrence Berkeley National Laboratory, 1 Cyclotron Road, Berkeley, CA 94720, USA}
\author{M.~Landriau}
\affiliation{Lawrence Berkeley National Laboratory, 1 Cyclotron Road, Berkeley, CA 94720, USA}
\author{L.~Le~Guillou}
\affiliation{Sorbonne Universit\'{e}, CNRS/IN2P3, Laboratoire de Physique Nucl\'{e}aire et de Hautes Energies (LPNHE), FR-75005 Paris, France}
\author{M.~E.~Levi}
\affiliation{Lawrence Berkeley National Laboratory, 1 Cyclotron Road, Berkeley, CA 94720, USA}
\author{M.~Manera}
\affiliation{Departament de F\'{i}sica, Serra H\'{u}nter, Universitat Aut\`{o}noma de Barcelona, 08193 Bellaterra (Barcelona), Spain}
\affiliation{Institut de F\'{i}sica d’Altes Energies (IFAE), The Barcelona Institute of Science and Technology, Campus UAB, 08193 Bellaterra Barcelona, Spain}
\author{A.~Meisner}
\affiliation{NSF NOIRLab, 950 N. Cherry Ave., Tucson, AZ 85719, USA}
\author{R.~Miquel}
\affiliation{Instituci\'{o} Catalana de Recerca i Estudis Avan\c{c}ats, Passeig de Llu\'{\i}s Companys, 23, 08010 Barcelona, Spain}
\affiliation{Institut de F\'{i}sica d’Altes Energies (IFAE), The Barcelona Institute of Science and Technology, Campus UAB, 08193 Bellaterra Barcelona, Spain}
\author{E.~Mueller}
\affiliation{Department of Physics and Astronomy, University of Sussex, Brighton BN1 9QH, U.K}
\author{A.~Muñoz-Gutiérrez}
\affiliation{Instituto de F\'{\i}sica, Universidad Nacional Aut\'{o}noma de M\'{e}xico,  Cd. de M\'{e}xico  C.P. 04510,  M\'{e}xico}
\author{A.~D.~Myers}
\affiliation{Department of Physics \& Astronomy, University  of Wyoming, 1000 E. University, Dept.~3905, Laramie, WY 82071, USA}
\author{J.~ A.~Newman}
\affiliation{Department of Physics \& Astronomy and Pittsburgh Particle Physics, Astrophysics, and Cosmology Center (PITT PACC), University of Pittsburgh, 3941 O'Hara Street, Pittsburgh, PA 15260, USA}
\author{J.~Nie}
\affiliation{National Astronomical Observatories, Chinese Academy of Sciences, A20 Datun Rd., Chaoyang District, Beijing, 100012, P.R. China}
\author{G.~Niz}
\affiliation{Departamento de F\'{i}sica, Universidad de Guanajuato - DCI, C.P. 37150, Leon, Guanajuato, M\'{e}xico}
\affiliation{Instituto Avanzado de Cosmolog\'{\i}a A.~C., San Marcos 11 - Atenas 202. Magdalena Contreras, 10720. Ciudad de M\'{e}xico, M\'{e}xico}
\author{N.~Palanque-Delabrouille}
\affiliation{IRFU, CEA, Universit\'{e} Paris-Saclay, F-91191 Gif-sur-Yvette, France}
\affiliation{Lawrence Berkeley National Laboratory, 1 Cyclotron Road, Berkeley, CA 94720, USA}
\author{W.~J.~Percival}
\affiliation{Department of Physics and Astronomy, University of Waterloo, 200 University Ave W, Waterloo, ON N2L 3G1, Canada}
\affiliation{Perimeter Institute for Theoretical Physics, 31 Caroline St. North, Waterloo, ON N2L 2Y5, Canada}
\affiliation{Waterloo Centre for Astrophysics, University of Waterloo, 200 University Ave W, Waterloo, ON N2L 3G1, Canada}
\author{C.~Poppett}
\affiliation{Lawrence Berkeley National Laboratory, 1 Cyclotron Road, Berkeley, CA 94720, USA}
\affiliation{Space Sciences Laboratory, University of California, Berkeley, 7 Gauss Way, Berkeley, CA  94720, USA}
\author{F.~Prada}
\affiliation{Instituto de Astrof\'{i}sica de Andaluc\'{i}a (CSIC), Glorieta de la Astronom\'{i}a, s/n, E-18008 Granada, Spain}
\author{C.~Ravoux}
\affiliation{Aix Marseille Univ, CNRS/IN2P3, CPPM, Marseille, France}
\affiliation{IRFU, CEA, Universit\'{e} Paris-Saclay, F-91191 Gif-sur-Yvette, France}
\affiliation{Universit\'{e} Clermont-Auvergne, CNRS, LPCA, 63000 Clermont-Ferrand, France}
\author{M.~Rezaie}
\affiliation{Department of Physics, Kansas State University, 116 Cardwell Hall, Manhattan, KS 66506, USA}
\author{A.~J.~Ross}
\affiliation{Center for Cosmology and AstroParticle Physics, The Ohio State University, 191 West Woodruff Avenue, Columbus, OH 43210, USA}
\affiliation{Department of Astronomy, The Ohio State University, 4055 McPherson Laboratory, 140 W 18th Avenue, Columbus, OH 43210, USA}
\author{G.~Rossi}
\affiliation{Department of Physics and Astronomy, Sejong University, Seoul, 143-747, Korea}
\author{E.~Sanchez}
\affiliation{CIEMAT, Avenida Complutense 40, E-28040 Madrid, Spain}
\author{E.~F.~Schlafly}
\affiliation{Space Telescope Science Institute, 3700 San Martin Drive, Baltimore, MD 21218, USA}
\author{D.~Schlegel}
\affiliation{Lawrence Berkeley National Laboratory, 1 Cyclotron Road, Berkeley, CA 94720, USA}
\author{H.~Seo}
\affiliation{Department of Physics \& Astronomy, Ohio University, Athens, OH 45701, USA}
\author{D.~Sprayberry}
\affiliation{NSF NOIRLab, 950 N. Cherry Ave., Tucson, AZ 85719, USA}
\author{T.~Tan}
\affiliation{IRFU, CEA, Universit\'{e} Paris-Saclay, F-91191 Gif-sur-Yvette, France}
\author{G.~Tarl\'{e}}
\affiliation{Department of Physics, University of Michigan, Ann Arbor, MI 48109, USA}
\author{B.~A.~Weaver}
\affiliation{NSF NOIRLab, 950 N. Cherry Ave., Tucson, AZ 85719, USA}
\author{H.~Zou}
\affiliation{National Astronomical Observatories, Chinese Academy of Sciences, A20 Datun Rd., Chaoyang District, Beijing, 100012, P.R. China}


\date{\today}

\begin{abstract}
The squeezed cross-bispectrum \bispeconed\ between the gravitational lensing in the Cosmic Microwave Background and the 1D \lya\ forest power spectrum can constrain bias parameters and break degeneracies between $\sigma_8$ and other cosmological parameters. 
We detect \bispeconed\ with $4.8\sigma$ significance at an effective redshift $z_\mathrm{eff}=2.4$ using Planck PR3 lensing map and over 280,000 quasar spectra from the Dark Energy Spectroscopic Instrument's first-year data. We test our measurement against metal contamination and foregrounds such as Galactic extinction and clusters of galaxies by deprojecting the thermal Sunyaev-Zeldovich effect. We compare our results to a tree-level perturbation theory calculation and find reasonable agreement between the model and measurement.
\end{abstract}


\maketitle

\section{\label{sec:intro}Introduction}
The observed large-scale structure of the universe was largely produced by the force of gravity acting on 
initial Gaussian density fluctuations. This non-linear evolution couples the originally linearly independent modes of the density field over time, and gives rise to a non-zero bispectrum and higher-order correlations. 
In the position-dependent power spectrum picture, the gravitational collapse will be faster in overdense regions due to the presence of more matter, which will then make the matter field clumpier and enhance the local small-scale power spectrum \citep{liSupersampleSignal2014, chiangPositiondependentPowerSpectrum2014}.

The \lya\ forest and Cosmic Microwave Background (CMB) lensing maps provide a rare opportunity to observe this effect as one needs a power spectrum and an accompanying large-scale density mode estimate at multiple locations in the Universe. The \lya\ forest technique maps out the matter field using the absorption lines in the quasar spectrum. The 1D power spectrum (\poned) of the \lya\ forest is most sensitive to small-scale physics and has been precisely measured in the redshift range of $2\lesssim z \lesssim 4$ \citep{palanque-delabrouilleOnedimensionalLyalphaForest2013, waltherNewPrecisionMeasurement2017, karacayliOptimal1DLy2022, ravouxFFTP1dEDR2023, karacayliOptimal1dDesiEdr2023}. The paths of the CMB photons are distorted by the intervening matter due to gravitational lensing, which was first detected in cross-correlations \citep{smithDetectionGravitationalLensingCmb2007, hirataCmbLssWeakLensing2008}. The lensing convergence $\kappa$ constructed from CMB temperature and polarization anisotropies corresponds to an integrated density field along the line of sight weighted by the lensing kernel $W_\kappa(\chi)$ \citep{bartelmannWeakLensing2017}. This kernel is broad, but peaks at $z\approx 2$, which overlaps with the redshift range of \lya\ \poned\ measurements.

The enhancement of \poned\ due to large-scale density fluctuations was proposed by \citet{zaldarriagaCorrelationsLyaForest2001}. The first measurement of the CMB lensing and \poned\ cross-bispectrum \bispeconed\ was reported by \citet{douxFirstDetectionCosmic2016} using about 87,000 SDSS-III/BOSS spectra \citep{dawsonBossSdssIii2013} at $5\sigma$. In this work, we measure \bispeconed\ using over 280,000 quasar spectra from the Dark Energy Spectroscopic Instrument's \citep[DESI,][]{leviDESIExperimentWhitepaper2013, desicollaborationDESIExperimentPart2016} first-year quasar sample. These quasar spectra are part of the future Data Release 1 \citep[DR1,][]{desiKp2DataRelease12024}. Key DESI science papers using DR1 already include the highest precision measurements of baryon acoustic oscillations (BAO) from galaxies and quasars \citep{desiKp4BaoGalaxies2024}, and from the \lya\ forest \citep{desiKp6BaoLya2024}, and the cosmological results from these BAO measurements \citep{desiKp7Cosmology2024}. To measure \poned, we apply the optimal estimator method \citep{mcdonaldLyUpalphaForest2006, karacayliOptimal1dDesiEdr2023}, which is robust against strong sky emission lines, defective CCD pixels, and low-signal-to-noise ratio (SNR) spectra. This estimator naturally provides a covariance matrix based on the large-scale correlations and the wavelength-dependent pipeline noise for each quasar and enables the optimal weighting for the measurement of \bispeconed.

Notable similar analyses include cross-correlations between the amplitude of \lya\ forest flux decrements and $\kappa$ \citep{vallinottoLensesForestCross2009, vallinottoCROSSCORRELATIONSLyaFOREST2011}, the galaxy-galaxy-$\kappa$ bispectrum with a $>20\sigma$ detection \citep{farrenDetectionCMBLensing2023}, and \lya\ forest and CMB temperature cross-correlations to study the Sunyaev-Zel’dovich effect \citep{croftLymanalphaForestCMBCrosscorrelation2006}.

The outline of the paper is as follows. We overview DESI quasar spectra and Planck lensing data in Section~\ref{sec:data}. We present the quasar continuum fitting algorithm, \bispeconed\ estimation, and measurement results in Section~\ref{sec:measurement}.
We develop a tree-level perturbation theory for \bispeconed, and present the best-fit results in Section~\ref{sec:theory}. We discuss measurement and modeling challenges in Section~\ref{sec:discuss}, and finally summarize in Section~\ref{sec:summary}.

\section{\label{sec:data}Data}
\subsection{DESI quasar spectra}
The DESI collaboration began a five-year survey in May 2021 to advance the understanding of the nature of dark energy through the most precise clustering measurements of galaxies, quasars, and the \lya\ forest ever obtained. DESI is mounted on the 4\ m Mayall telescope and can obtain 5000 spectra in each observation \citep{desicollaborationDESIExperimentPart2016b, silberRoboticMultiobjectFocal2023}. It has ten, identical spectrographs that are in a climate-controlled enclosure to minimize instrumental systematic errors \citep{abareshiOverviewInstrumentationDark2022, guySpectroscopicDataProcessingPipeline2022}. DESI target selection was based on the photometry from the Legacy Imaging Surveys \citep{deyOverviewDESILegacy2019} and the Wide-field Infrared Explorer \citep{wrightWideFieldWISE2010}, and is described in detail in \citet{myersTargetSelectionPipelineDESI2022}. The collaboration refined the target selection algorithms during the Survey Validation \citep{surveyValidation2023} period in early 2021 with a significant visual inspection effort \citep{alexanderDESISVVIQSO2022}. The preliminary quasar target sample is presented in \citet{yechePreliminaryTargetQSODESI2020} and the final quasar target selection in \citet{chaussidonTargetSelectionDESIQSO2022}.

We use the DR1 quasar observations in this analysis. This sample has over 1.5 million quasars observed between December 2020 and June 2022 \citep{ashleyDesiLssCatalog2024}. Nearly 450,000 of these quasars are at $z > 2.1$, and therefore the DESI spectra include the \lya\ forest region.  At the end of its five-year mission, DESI is expected to collect approximately 800,000 \lya\ quasars ($z > 2.1$) and have twice as much exposure time as DR1 per quasar spectrum on average, which could improve the \bispeconed\ covariance matrix by a factor of 3.5.

Broad absorption line (BAL) features are identified using an algorithm similar to the one presented by \citet{guoClassificationBALs2019}, except that it does not use the Convolutional Neural Network (CNN) classifier. A detailed study of these features will be presented by \citet[][in preparation]{martiniDesiBalY12024}. 18.5\% of our DR1 quasars have BAL features. We mask the wavelength ranges of S~\textsc{iv}, P~\textsc{v}, C~\textsc{iii}, \lya, N~\textsc{v}, and Si~\textsc{iv} ions based on the velocity ranges of the absorption troughs observed for the C~\textsc{iv} BAL features outside of the forest region.

Damped \lya\ systems (DLA) are identified using a Convolutional Neural Network (CNN) \citep{mingfengDLAGP2021, wangDeepLearningDESIDLA2022}. We mask all DLAs ($\log N_\mathrm{HI}>20.3$) except the ones with confidence level less than 0.3 in quasars with $\overline{\mathrm{SNR}}<3$, where the average signal-to-noise ratio $\overline{\mathrm{SNR}}$ is calculated between 1420--1480~\AA\ in quasar's rest frame \citep{karacayliOptimal1dDesiEdr2023}. There are 88,858 DLAs observed in 73,513 quasars in our catalog. We mask the regions where the model DLA transmission profile is below 80\% and correct the damping wings at larger transmission values based on the same model profile \citep{karacayliOptimal1dDesiEdr2023}. These systems and also neutral hydrogen systems with $\log N_{H \textsc{i}} \gtrsim 19$ are expected to be correlated with CMB lensing, which we further address in Section~\ref{sec:measurement}.

\subsection{Gravitional lensing}
We use the minimum-variance (MV) CMB lensing convergence map $\hat{\kappa}_{LM}$ of Planck 2018 PR3\footnote{\url{https://pla.esac.esa.int/pla/\#cosmology}}\cite{collaborationPlanck2018OGravitationalLensing2020} (we discuss the performance of PR4 in Section~\ref{sec:discuss}). This map is stored in $N_\mathrm{side}=2048$ \texttt{HEALPix} pixelization \citep{healpix} with modes up to $L_\mathrm{max}=4096$ available. We construct the Wiener filter $W_\mathrm{WF} = C_L / (C_L + N_L)$ from the noise $N_L$ and signal $C_L$ power spectrum of $\hat{\kappa}_{LM}$.

In our baseline analysis, we apply the Wiener filter to $\hat{\kappa}$ and keep all the angular modes ($0\leq L \leq 4096$). We perform one variation where we limit modes to the recommended conservative region of $8\leq L \leq 400$ by setting the other $L$ modes to zero. At $z=2.4$, the angular mode $L=400$ corresponds to  comoving wavenumber $k=L/\chi(2.4)\approx 0.1~$\hmpc. We then calculate $\kappa_q$ for each quasar in our catalog by bilinear interpolation.


\section{\label{sec:measurement}Measurement}
\subsection{\label{subsec:contfit}Continuum fitting}
We use the standardized continuum fitting algorithm that was developed over the years and has been applied to both $\xi_\mathrm{3D}$ and \poned\ measurements \citep{bourbouxCompletedSDSSIVExtended2020, karacayliOptimal1dDesiEdr2023}. We summarize the algorithm below and refer the reader to \citet{bourbouxCompletedSDSSIVExtended2020}, \citet{ramirezperezLyaCatalogDesiEdr2023} and \citet{karacayliOptimal1dDesiEdr2023} for a detailed description.

In this continuum fitting framework, the definition of the quasar continuum absorbs the mean transmission of the IGM $\overline{F}(z)$, such that the quasar ``continuum" $\overline{F} C_q(\lambda_\mathrm{RF})$ is given by
\begin{align}
    \overline{F}C_q(\lambda_\mathrm{RF}) &= \overline{C}(\lambda_\mathrm{RF}) \left( a_q + b_q \Lambda \right) \\
    \Lambda &= \frac{\log\lambda_\mathrm{RF} - \log\lambda_\mathrm{RF}^{(1)}}{\log\lambda_\mathrm{RF}^{(2)} - \log\lambda_\mathrm{RF}^{(1)}},
\end{align}
where $\lambda_\mathrm{RF}$ is the wavelength in the quasar's rest frame, $\lambda_\mathrm{RF}^{(1, 2)}$ are the minimum and maximum wavelengths considered in the calculation, $\overline{C}(\lambda_\mathrm{RF})$ is the global mean continuum, and finally $a_q$ and $b_q$ are two quasar diversity parameters. Note that these parameters do not only fit for intrinsic quasar diversity such as brightness, but also for the IGM mean transmission. Given these definitions, transmitted flux fluctuations are given by
\begin{equation}
    \delta_F^{(q)}(\lambda) = \frac{f_q(\lambda)}{\overline{F}C_q(\lambda_\mathrm{RF})} - 1,
\end{equation}
where $\lambda=(1+z_q)\lambda_\mathrm{RF}$\ is the observed wavelength and $f_q(\lambda)$ is the observed flux. 

This continuum fitting algorithm assigns each pixel a variance $\sigma_{q}^2(\lambda)$:
\begin{equation}
    \label{eq:sigma2_cfit}\sigma_{q}^2(\lambda) = \eta(\lambda) \sigma^2_\mathrm{pipe}(\lambda) + \sigma^2_\mathrm{LSS}(\lambda) (\overline{F}C_q)^2(\lambda),
\end{equation}
based on the observed variance statistics of $\delta_F$, which includes a pipeline noise correction $\eta(\lambda)$ term as a scaling of the pipeline variance estimates and an additive large-scale structure variance term $\sigma^2_\mathrm{LSS}(\lambda)$. In the \lya\ forest region of the quasar spectrum, the $\sigma^2_\mathrm{LSS}$ term is dominated by actual large-scale fluctuations. However, outside that region, it mostly reflects the additive errors made by the pipeline in the noise estimation since large-scale fluctuations due to metal absorption contribute little to this term. 

In \citet{karacayliOptimal1dDesiEdr2023}, we found that the noise calibration errors originate at the CCD level, and we proposed data splits based on spectrograph and CCD amplifier location to mitigate these errors. Even though the cross-correlation signal is not biased due to noise calibration errors, a noise recalibration step moderately improves the significance of our cross-correlation detection. Regular analyses use the spectra in \emph{healpix} grouping in which the multiple, different exposures of the same object are co-added. To calculate and propagate this CCD-amplifier-dependent correction, we use the \emph{tile} grouping of DESI observations, where a tile is a fixed pointing of the telescope with fixed fiber assignments to specific targets \citep{earlyDataRelease2023}. This grouping guarantees that the same quasar will be observed by the same spectrograph at the same (approximate) location on the CCD if it is observed in the same tile. We split the data into 20 subsets based on spectrograph and CCD amplifier region, and calculate the $\eta$ and $\sigma^2_\mathrm{LSS}$ values between 1600--1800~\AA\ in the quasar's rest frame for each subset. This range excludes the dominant Si~\textsc{iv} and C~\textsc{iv} systems, but is still affected by weak Mg~\textsc{ii} absorption.

We then perform the continuum fitting for each subset of quasar spectra using \texttt{qsonic}\footnote{\url{https://qsonic.readthedocs.io/en/stable/}}~\citep{Karacayli_QSOnic_fast_quasar_2024} while correcting each CCD amplifier region's noise estimates using the $\eta$ and $\sigma^2_\mathrm{LSS}$ values obtained above. We define the \lya\ forest region to be between 1045~\AA\ $< \lambda_\mathrm{RF} <$ 1185~\AA, and a sideband region to be 1268~\AA\ $< \lambda_\mathrm{RF} <$ 1380~\AA. We use this sideband region to quantify the possible metal contributions to the cross-correlation. We limit the observed wavelength coverage to  3600~\AA\ $< \lambda <$ 5350~\AA, which only resides in the blue channel. The wavelengths contaminated by sky lines are naturally down-weighted by the pipeline, but we mask certain particularly strong lines due to difficulties in modeling\footnote{\url{https://github.com/corentinravoux/p1desi/blob/main/etc/skylines/list_mask_p1d_DESI_EDR.txt}}. For the \lya\ forest region only, we use the fiducial mean flux of \citet{beckerMeanFlux2013} in \texttt{qsonic} to alleviate the coupling between $\overline{F}$ and quasar diversity parameters.

We select quasars with an average SNR greater than one at wavelengths greater than the \lya\ emission line of the quasar, and with an average SNR greater than 0.3 in the forest region. After these steps, we have 299,781 quasars that satisfy the SNR cuts and that produce valid continuum fits. We then use the 278,098 quasars that overlap with the Planck lensing map.

Since the quadratic estimator is essentially a weighted average, low-SNR spectra are naturally down-weighted, so these cuts aim to remove possibly non-quasar objects from our input. For example, an SNR cut of 2 in the forest region keeps the top 34\% of spectra, but increases the covariance matrix by about only 20\%.

\subsection{Power spectrum and cross bispectrum estimation}
The quadratic maximum likelihood estimator (QMLE) gives a power spectrum $\bm{p}_q$ and its covariance matrix $\mathbf{C}_q$ for each quasar $q$ \cite{karacayliOptimal1DLy2020, karacayliOptimal1dDesiEdr2023}. This covariance matrix $\mathbf{C}_q$ includes contributions from the pipeline noise estimates and the signal contribution based on a fiducial power spectrum (see Appendix~\ref{app:qmle} for a summary). Using these matrices as inverse weights, we build the following bispectrum estimator:
\begin{equation}
    \label{eq:bispec_estimator}\hat{B}_a = N_{ab} \sum_q W^q_{bc} \left(p^q_{c} - \hat{\overline{{p}}}_c \right) \left(\kappa^q - \hat{\overline{{\kappa}}}_b \right),
\end{equation}
where indices $a, b, c$ correspond to $(z, k)$ bins; $b, c$ are summed over per Einstein notation; and weights are given by $\mathbf{W}_q = \mathbf{C}_\mathrm{tot}\mathbf{C}^{-1}_q$ and $\mathbf{C}_\mathrm{tot} \equiv \sum_q \mathbf{C}_q$ such that $\sum_q \mathbf{W}_q = \mathbf{I}$. The mean power spectrum and $\kappa$ are estimated using these weights:
\begin{align}
    \bm{\hat{\overline{{p}}}} &= \sum_q \textbf{W}_q \bm{p}_q, \\
    \hat{\overline{{\kappa}}}_a &= \sum_{q} \kappa_q \sum_b W^q_{ab}.
\end{align}
The sample bias correction term $\mathbf{N}$ is given by
\begin{equation}
    \label{eq:sample_bias}N^{-1}_{ab} = \delta^K_{ab} - \sum_q \left(\mathbf{W}^2_q\right)_{ab}.
\end{equation}
We provide a derivation of the sample bias correction in Appendix~\ref{app:derive_N}

We measure \poned\ in 35 linearly-spaced $k$ bins with $\Delta k = 6\times10^{-4}~$\skm\ and in 4 redshift bins with $\Delta z=0.4$ starting at $z=2.0$. Unfortunately, the detection significance in each $z$ bin is low, so we average all $z$ bins using the total Fisher matrix ($\mathbf{F}_\mathrm{tot} = \mathbf{C}^{-1}_\mathrm{tot}$) to improve our measurement. The redshift-averaged estimates are given by a linear operation $\hat{\bar{B}}_k = T_{ka} \hat{B}_a$, where
\begin{equation}
    T_{k a} \propto \sum_{z} F^\mathrm{tot}_{(z, k), a},
\end{equation}
such that each row of $T$ is normalized, $\sum_a T_{k a} = 1$. Note that this choice correctly takes the correlations between redshift bins into account in averaging.

The estimator in Eq.~\ref{eq:bispec_estimator} can be straightforwardly extended to $\theta$ separations between \poned\ and $\kappa$ by calculating $\kappa_q$ as a mean inside the ring around the quasar for an angular width $\Delta \theta$. The resulting 2D bispectrum $\zeta(k, \theta)$ is in Fourier space in the radial direction and in real space in the angular direction. Since the Wiener filter suppresses angular fluctuations below $\theta < 0.45^\circ$ $(L>400)$, it means that $B_{\kappa, \mathrm{Ly}\alpha} \approx \zeta(k, \theta \lesssim 0.45^\circ)$ and also $\zeta(k, \theta)$ will be highly correlated between angular bins separated by the same angle. Furthermore, $\theta=0.45^\circ$ corresponds to $r_\perp = 31~$\mpch separation, and $\zeta(k, \theta)$ decreases to non-detectable levels by $r_\perp = 70~$\mpch. To observe the angular dependence more finely, we use slightly narrower $\theta$ bins that are linearly-spaced and centered at $0.2^\circ, 0.4^\circ, 0.6^\circ$, and $0.8^\circ$ with $\Delta\theta=0.2^\circ$.

We estimate the covariance matrix using the non-parametric bootstrap method \citep{efronBootstrap1979}. To account for the angular correlations, we group \bispeconed\ measurements into continuous regions on the sky using HEALPix pixelization scheme \citep{healpix}. DESI organizes spectra using $N_\mathrm{side}=64$, which corresponds to an angular resolution of $0.92^\circ$. This is larger than the Wiener filter scale, and captures the large angular scale correlations in bootstrap realizations. There are 12,803 healpixels. We randomly assign an observation frequency for each healpix based on a Poisson distribution with a mean of one \citep{hanleyNonParametricPoissonBootstrap2006, chamandy2012estimating}. Using the healpix number as the seed in random number generation generates correlated bootstrap realization between angular bins. To speed up these repeated runs, we do not recalculate the sample bias correction term, and use the original sample's value for each realization. We calculate the covariance matrix with 50,000 realizations.

\subsection{Software}
Our quadratic estimator\footnote{\url{https://github.com/p-slash/lyspeq}} is written in \texttt{c++}.
It depends on \texttt{CBLAS} and \texttt{LAPACKE} routines for matrix/vector operations, \texttt{GSL}\footnote{\url{https://www.gnu.org/software/gsl}} for certain scientific calculations \citep{GSL}, \texttt{FFTW3}\footnote{\url{https://fftw.org}} for fast Fourier transforms \citep{FFTW05}; and uses the Message Passing Interface (MPI) standard\footnote{\url{https://www.mpich.org}} to parallelize tasks.
The quasar spectra are organized with the \texttt{HEALPix} \citep{healpix} scheme on the sky.
We use the following commonly-used software in \texttt{python} analysis: \texttt{astropy}\footnote{\url{https://www.astropy.org}}
a community-developed core \texttt{python} package for Astronomy \citep{astropy:2013, astropy:2018, astropy:2022},
\texttt{numpy}\footnote{\url{https://numpy.org}}
an open source project aiming to enable numerical computing with \texttt{python} \citep{numpy},
\texttt{scipy}\footnote{\url{https://scipy.org}}
an open-source project with algorithms for scientific computing,
\texttt{healpy} to interface with \texttt{HEALPix} in \texttt{python} \citep{healpy},
\texttt{numba}\footnote{\url{https://numba.pydata.org}}
an open source just-in-time (JIT) compiler that translates a subset of \texttt{python} and \texttt{numpy} code into fast machine code,
\texttt{mpi4py}\footnote{\url{https://mpi4py.readthedocs.io}}
which provides \texttt{python} bindings for the MPI standard \citep{mpi4py}.
Finally, we make plots using
\texttt{matplotlib}\footnote{\url{https://matplotlib.org}}
a comprehensive library for creating static, animated, and interactive visualizations in \texttt{python}
\citep{matplotlib}.

\subsection{\label{subsec:results}Results}
\begin{figure}
    \centering
    \includegraphics[width=\columnwidth]{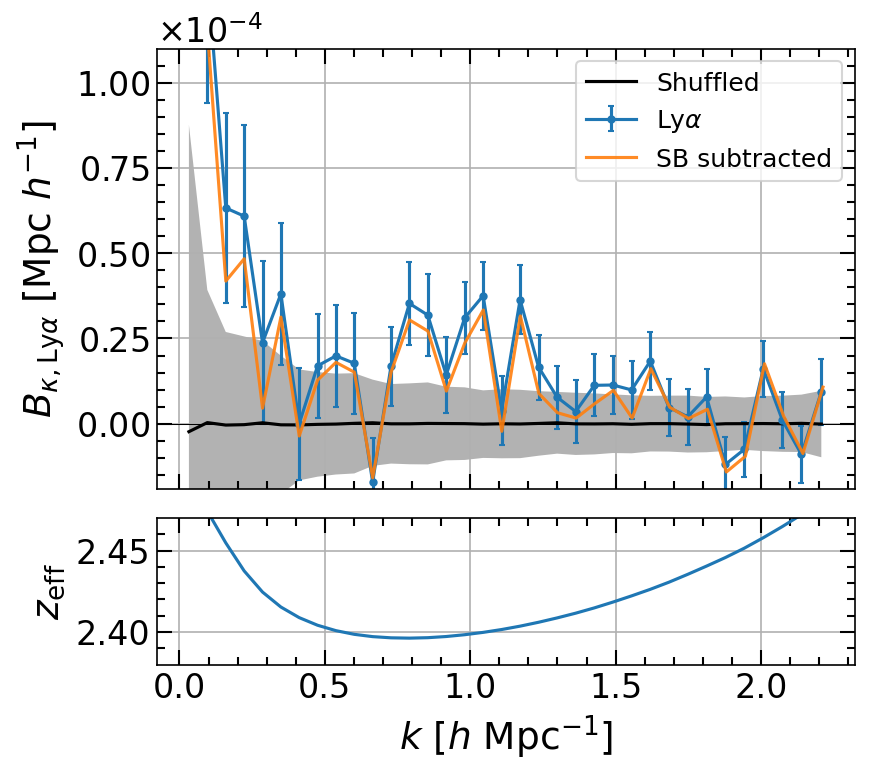}
    \caption{\label{fig:b1d_allmodes}We detect a redshift averaged \bispeconed\ ({\it blue circles}) at $4.8\sigma$ significance in our baseline analysis. The shuffle test yields a non-detection ({\it black line}). The sideband bispectrum subtraction to remove metal contamination ({\it orange line}) decreases the measurement values by around 15\% on average. The bottom panel shows the effective redshift of each $k$ bin, which shows a mild trend between 2.40--2.45.}
\end{figure}

\begin{figure}
    \centering
    \includegraphics[width=\columnwidth]{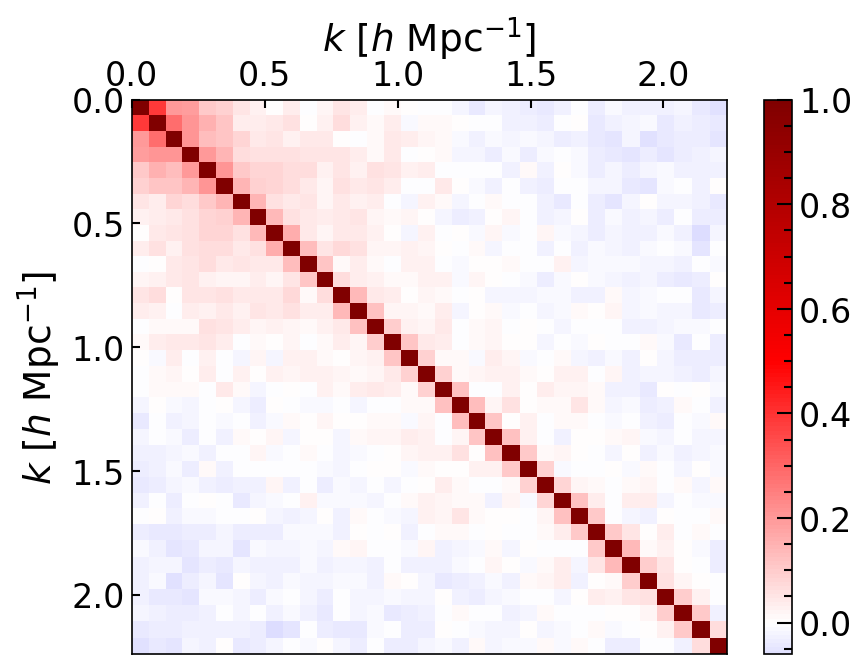}
    \caption{\label{fig:b1d_cov}Correlation matrix based on the bootstrap covariance matrix of \bispeconed. Adjacent $k$ bins are correlated by 10--20\%, whereas the correlations between large-scale modes (small $k$) range between 20--40\%. We do not find any correlations between \poned\ and \bispeconed.}
\end{figure}

Figure~\ref{fig:b1d_allmodes} shows our average \bispeconed\ measurement at an effective redshift of $z=2.44$.
The effective redshift per $k$ bin shows a mild trend between $z=2.40-2.45$ as can be seen in the bottom panel of Figure~\ref{fig:b1d_allmodes}. The black line is the average result of the shuffled $\kappa$ test, where we randomly assign $\kappa$ to each quasar, and the grey-shaded region is the amplitude of the scatter between all shuffled bispectra. Figure~\ref{fig:b1d_cov} shows the correlation matrix based on the bootstrap covariance matrix, $r_{ij} = C_{ij} / \sqrt{C_{ii} C_{jj}}$.
We find that the large-scale modes, $k < 0.15~$\hmpc, are highly correlated (up to 40\%) and adjacent $k$ bins are 15\% correlated on average. We also calculate the cross-covariance between \poned\ and \bispeconed, and find no correlations $(<1\%)$ between them at our current precision.

Using this covariance matrix, we detect a 1D \lya\ forest-$\kappa$ bispectrum signal by $4.8\sigma$ when we keep all the angular modes in the lensing map, and by $3.8\sigma$ when we limit them to the conservative region of $8\leq L \leq 400$. As we show in Section~\ref{sec:theory}, the signal itself depends on the Wiener filter applied to the lensing map, and therefore this decrease in detection significance can be attributed to both loss in the number of modes and loss in the correlation signal itself.

\begin{figure}
    \centering
    \includegraphics[width=\columnwidth]{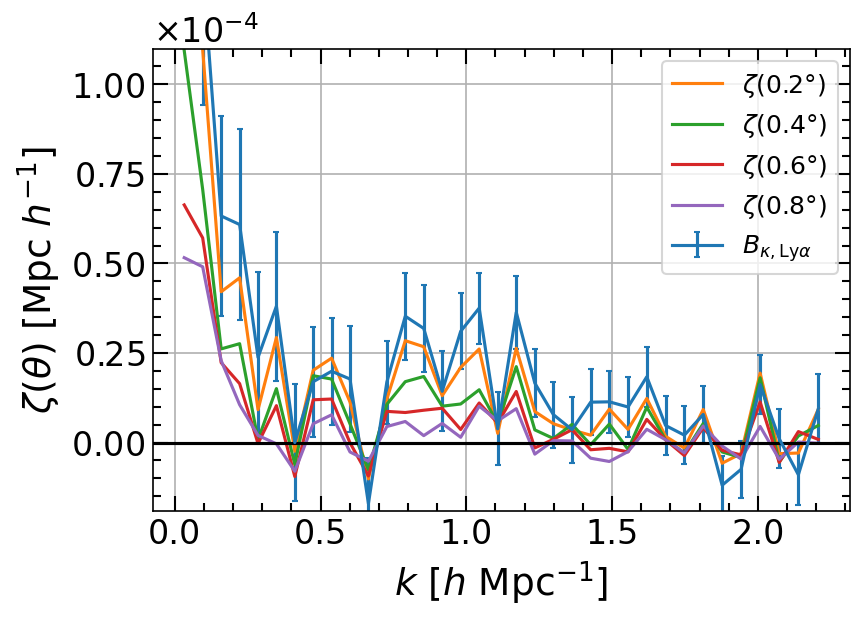}
    \caption{2D bispectrum $\zeta(k, \theta)$ for four $\theta$ bins and $B_{\kappa, \mathrm{Ly}\alpha}(k) = \zeta(k, \theta=0^\circ)$ bin. The amplitude of $\zeta(k, \theta)$ decreases as the angular separation increases as expected. The detection significance in each angular bin is $4.1\sigma, 2.1\sigma, 1.1\sigma,$ and no detection respectively, and the detection significance of the combined data vector is only $2.7\sigma$ due to high correlations.}
    \label{fig:bispec_angular_bins}
\end{figure}

\begin{figure}
    \centering
    \includegraphics[width=\columnwidth]{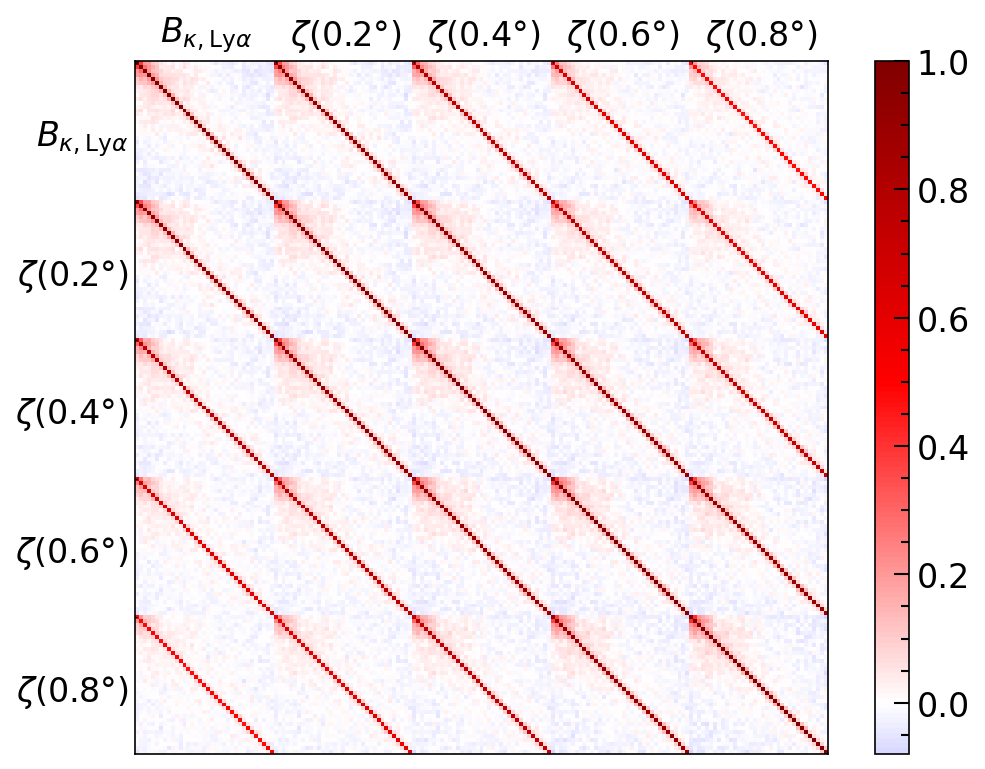}
    \caption{Correlation matrix of the combined data vector. The correlation coefficient decreases from over 90\% between the adjacent angular bins, to 50\% between \bispeconed\ and $\zeta(\theta=0.8^\circ)$ bins.}
    \label{fig:bispec_angular_cov}
\end{figure}

Figure~\ref{fig:bispec_angular_bins} shows the 2D bispectrum $\zeta(k, \theta)$ for four $\theta$ bins and $B_{\kappa, \mathrm{Ly}\alpha}(k)$ which is the same as $\zeta(k, \theta=0^\circ)$. The correlations between \poned\ and $\kappa$ diminish at larger separations as expected. Due to the Wiener filter suppression of angular fluctuations below $\Delta \theta \lesssim 0.45^\circ$ scale, these additional measurements are highly correlated, which can be seen in Figure~\ref{fig:bispec_angular_cov}. The recurrent fluctuations at the same $k$ values between angular bins are most likely correlated. The correlation coefficient decreases from over 90\% between the adjacent angular bins, to 50\% between \bispeconed\ and $\zeta(\theta=0.8^\circ)$ bins. The detection significance in each individual angular bin is as follows: $4.1\sigma, 2.1\sigma, 1.1\sigma,$ and no detection respectively. Note that at the largest angular separation, the correlation naturally goes to zero and results in a ``non-detection". Since the combined data vector is highly correlated, its detection significance is only $2.7\sigma$. The reduced significance might come as surprising, but it is due to correlations. If we had ignored correlations between angular bins, the significance would have been $5.9\sigma$.
Because of these complications, we focus on \bispeconed\ for the rest of the paper.

The detection significance of $4.8\sigma$ could be attributed to several possible absorption sources besides the diffuse intergalactic medium gas. These sources are high-column density (HCD) systems and metals. To isolate the \lya\ forest contribution, we further refine our bispectrum measurement using \poned\ measurement techniques. We note that \citet{douxFirstDetectionCosmic2016} has no such refinements, therefore $4.8\sigma$ remains the comparable quantity between the two studies.

First, the continuum marginalization affects the largest scales in the bispectrum, similar to \poned. These modes are further contaminated by HCDs, which include unidentified DLAs and any neutral hydrogen system with $\log N_{H \textsc{i}} \gtrsim 19$. These effects add large power to small $k$ bins and strongly correlate them, as seen in Figures~\ref{fig:b1d_allmodes} and~\ref{fig:b1d_cov}. We remove the first two $k$ bins to be conservative for these reasons. This has a relatively minor effect of reducing the detection significance to $4.5\sigma$.

Second, we estimate the metal contamination in the cross-correlation by estimating the bispectrum in the sideband (SB) region. We find that all the $k$ modes are highly correlated due to the weak nature of the signal, correlated fluctuations sourced by systematics in this region, and signal originating from line profiles \citep{karacayliFrameworkMetals2023}. Hence, we detect a metal-$\kappa$ cross-correlation at a weaker significance (around $1.6\sigma$). We subtract this $\kappa-$SB bispectrum from the $\kappa-$\lya\ bispectrum to statistically remove the metal contamination (orange line in Figure~\ref{fig:b1d_allmodes}), and add the bootstrap covariance matrices such that $\mathbf{C}_\mathrm{new} = \mathbf{C}_\mathrm{Lya} + \mathbf{C}_\mathrm{SB}$. This step individually reduces the detection significance to $3.0\sigma$. Further removing the first two $k$ bins reduces our detection significance to $2.7\sigma$.

We emphasize that that detection of absoption$-\kappa$ correlation remains $4.8\sigma$, but allowing for metal and HCD contributions means detection of \lya\ alone is $2.7\sigma$.

\section{\label{sec:theory}Theory}
In this section, we present a standard perturbation theory framework for \bispeconed, evaluate it at the tree level, and provide best-fit values for the \lya\ forest parameters. We also revisit and correctly incorporate the Wiener filter of the CMB lensing map into the position-dependent formalism.

An integral of the density field in the line of sight gives the lensing convergence $\kappa$:
\begin{align}
    \kappa =& \int W_\kappa(\chi) \delta_m(\chi) \mathrm{d}\chi, \\
    W_\kappa(\chi) =& \frac{3 H_0^2 \Omega_{m,0}}{2c^2} \frac{ (\chi_\mathrm{CMB} - \chi) \chi}{a\chi_\mathrm{CMB}},
\end{align}
where $\chi$ is the comoving distance and $\chi_\mathrm{CMB}$ is the comoving distance to the CMB source plane at $z_\mathrm{CMB}=1100$ \cite{bartelmannWeakLensing2017}. To obtain the cross-correlations between \poned\ and $\kappa$ at redshift $z_*$, we employ the Limber approximation and limit the density field contributions to $\kappa$ to the redshift range such that the lensing kernel is approximately constant $W_\kappa(z) \approx W_\kappa(z_*)$ (see Appendix~\ref{app:theory_derivation} for the expressions without the Limber approximation). 
\begin{equation}
    \kappa(\bm{x}) = W_\kappa(z_*) \int \mathrm{d}\chi\, \delta_m(\chi, \bm{x}),
\end{equation}
where $\bm{x}$ is the 2D vector perpendicular to the line of sight. Wiener filtering smoothes the field on this surface:
\begin{equation}
    \delta^\mathrm{WF}_m(\chi, \bm{x}) = \int \frac{d^2\bm{p}_\bot}{(2\pi)^2} e^{i\bm{p}_\bot \cdot \bm{x}} W_\mathrm{WF}(p_\bot \chi_*) \Tilde{\delta}_m(\chi, \bm{p}_\bot),
\end{equation}
where we used the flat-sky approximation $l=p_\bot \chi_*$ and defined $\chi_* \equiv \chi(z_*)$.

We now begin the 1D \lya$-\kappa$ bispectrum calculation, which we define as follows:
\begin{align}
    \langle \delta_F(k, \bm{x}) \delta_F(k', \bm{x}) \kappa(\bm{x}) \rangle \equiv 2\pi \delta_D(k+k') B^\mathrm{1D}_{FF\kappa}(k) \\
    = W_\kappa(z_*) \int \mathrm{d}\chi\, \langle \delta_F(k, \bm{x}) \delta_F(k', \bm{x}) \delta^\mathrm{WF}_m(\chi, \bm{x}) \rangle.
\end{align}
The density fields can be written in terms of 3D Fourier transforms:
\begin{align}
    \delta_F(k, \bm{x}) &= \int \frac{d^2 \bm{q}_\bot}{(2\pi)^2} e^{i \bm{q}_\bot \cdot \bm{x}} \Tilde{\delta}_F(\bm{q}_\bot, k), \\
    \delta^\mathrm{WF}_m(\chi, \bm{x}) &= \int \frac{d^3\bm{p}}{(2\pi)^3} e^{i \bm{p} \cdot (\bm{x}, \chi)} W_\mathrm{WF}(p_\bot \chi_*) \Tilde{\delta}_m(\bm{p}),
\end{align}
and then the terms in brackets will yield the 3D bispectrum between \lya\ forest and matter fields.
\begin{equation}
    \langle \Tilde{\delta}_F(\bm{q}) \Tilde{\delta}_F(\bm{q}') \Tilde{\delta}_m(\bm{p}) \rangle = (2\pi)^3 \delta^D_{\bm{q}\bm{q}'\bm{p}} B^\mathrm{3D}_{FFm}(\bm{q}, \bm{p})
\end{equation}
Note that the integration over $d\chi dp_z$ sets $p_z=0$ and yields $2\pi \delta_D(k+k')$. The ensuing calculus yields the following expression for the 1D $\kappa-$\lya\ bispectrum:
\begin{align}
    \label{eq:b1dffm}B^\mathrm{1D}_{FF\kappa}(k, z) &= W_\kappa(z) \int \frac{d^2 \bm{q}_\bot d^2 \bm{p}_\bot}{(2\pi)^4} \nonumber \\
    &\times B^\mathrm{3D}_{FFm}(\bm{q}, \bm{q}', \bm{p}) W_\mathrm{WF}(p_\bot \chi_*),
\end{align}
where $\bm{q} = (\bm{q}_\bot, k)$, $\bm{p} = (\bm{p}_\bot, 0)$ and $\bm{q}' = (-\bm{q}_\bot - \bm{p}_\bot, -k)$. We found the integration over $\bm{p}_\bot$ to be more robust against the divergences in the tree-level bispectrum expansion. We present a cross-bispectrum expression including angular modes in Appendix~\ref{app:theory_derivation}.

These expressions are in units of \mpch, whereas \lya\ \poned\ is conventionally calculated in velocity units. The conversion factor between these two units is dependent on redshift and cosmology:
\begin{equation}
    \label{eq:mpc2kms}1~\mathrm{Mpc}~h^{-1} = 100 \frac{E(z)}{1+z} ~\mathrm{km~s}^{-1}.
\end{equation}

Accurate analytical modeling of $B^\mathrm{3D}_{FFm}$ is a difficult task not only because of the non-linear gravitational evolution but also because of nuisance parameters needed to describe the non-linearities in the \lya\ forest. The effective field theory for the \lya\ forest is a promising avenue to incorporate all the nuisance parameters \citep{ivanovEffectiveLya2024}, but is outside the scope of this work. Instead, we are going to limit our analysis to a simple large-scale structure bias model for $\delta_F$ with some pressure smoothing and thermal broadening such that
\begin{equation}
    \tilde{\delta}_F(\bm{q}) = b_F (1 + \beta_F \mu^2) \tilde{\delta}_m(\bm{q}) e^{-q^2 / k_p^2 - q_z^2 \sigma_{th}^2 / 2}, 
\end{equation}
where $k_p$ is the pressure smoothing/filtering scale \citep{gnedinProbingUniverseLya1998}, and $\sigma_{th}= \sqrt{k_B T/m_p}$ is the thermal broadening scale where we reparameterize temperature into a power law such that $T \rightarrow T_0 10^{\log T - 4}$ at $T_0=10^4~$K. Since the bispectrum needs a non-linear treatment even to first order and our measurement is at a modest $3.0\sigma$ detection significance, we ignore non-linear enhancement terms and additional angular dependencies proposed in the fitting functions in the literature \citep{mcdonaldPredictingLyaPower2003, arinyoNonLinearPowerLya2015}. We note that $\sigma_{th}$ is in velocity units and needs to be converted to \mpch\ units using Eq.~\ref{eq:mpc2kms}. We then calculate the matter bispectrum using  tree-level perturbation theory up to second-order
\begin{equation}
    B^\mathrm{3D}_{mmm} = 2 F_2(\bm{q}, \bm{p}) P_L(q) P_L(p) + \mathrm{two\ cyc.\ terms},
\end{equation}
where $P_L$ is the linear matter power spectrum and the $F_2$ kernel is \cite{jainSecondOrderPt1994}
\begin{equation}
	F_{2}(\bm{q}, \bm{p}) = \frac{5}{7} + \frac{\bm{q} \cdot \bm{p}}{2}\left[ \frac{1}{q^2} +\frac{1}{p^2} \right] + \frac{2}{7} \frac{(\bm{q} \cdot \bm{p})^{2}}{q^{2}p^{2}}.
\end{equation}
Symmetry allows us to write the integral in Eq.~\ref{eq:b1dffm} as follows:
\begin{equation}
    \int \frac{d^2 \bm{q}_\bot d^2 \bm{p}_\bot}{(2\pi)^4} \rightarrow \int \frac{p^2_\bot d \ln p_\bot}{2\pi} \int \frac{q^2_\bot d \ln q_\bot d\phi}{(2\pi)^2}.
\end{equation}
Furthermore, since the integrand only depends on $w \equiv \cos\phi$, we integrate the angular part by using the Chebyshev–Gauss quadrature method:
\begin{equation}
    \int_0^{2\pi} d\phi \rightarrow 2 \int_{-1}^{1} \frac{dw}{\sqrt{1 - w^2}}.
\end{equation}

There are two major shortcomings of our model. First, the second-order bias terms such as $b_2$ of the \lya\ forest will contribute to tree-level expressions at the leading order. Second, additional non-linear enhancements and angular-dependent effects are expected to contribute to the model similar to \poned. As noted above in the text, the effective field theory for the \lya\ forest can incorporate all these terms.

\subsection{Fit to measurement}
A full cosmological analysis is unlikely to be successful given the current error bars on \bispeconed\ and is therefore outside the scope of our paper. Instead, we fix the cosmological parameters to Planck 2018 values \citep{collaborationPlanck2018Results2020}: $\Omega_b h^2=0.02242, \Omega_c h^2=0.11934, h=0.6766, n_s=0.9665, \ln(10^{10} A_s) = 3.044$. Since \bispeconed\ integrates over the angular modes, it is less sensitive to the redshift space distortion parameter $\beta_F$. So, we also fix $\beta_F=1.67$ from the \lya\ 3D correlation function analysis \citep{bourbouxCompletedSDSSIVExtended2020}. We add Gaussian priors of $k_p = 7 \pm 30~$\hmpc\ and $\log T = 4 \pm 2$ \citep{waltherNewConstraintsIGM2019, gaikwadConsistentRobustMeasurement2020, villasenorWarmDarkMatter2023}. We then fit this model to our \bispeconed\  measurement that has had the side band metal contamination and the first two $k$ bins removed. We calculate the linear matter power spectrum using \texttt{CosmoPower} \citep{spuriomanciniCosmopower2022} and minimize $\chi^2$ using \texttt{iminuit} \citep{jamesMinuit1975, iminuit}.

\begin{figure}
    \centering
    \includegraphics[width=\columnwidth]{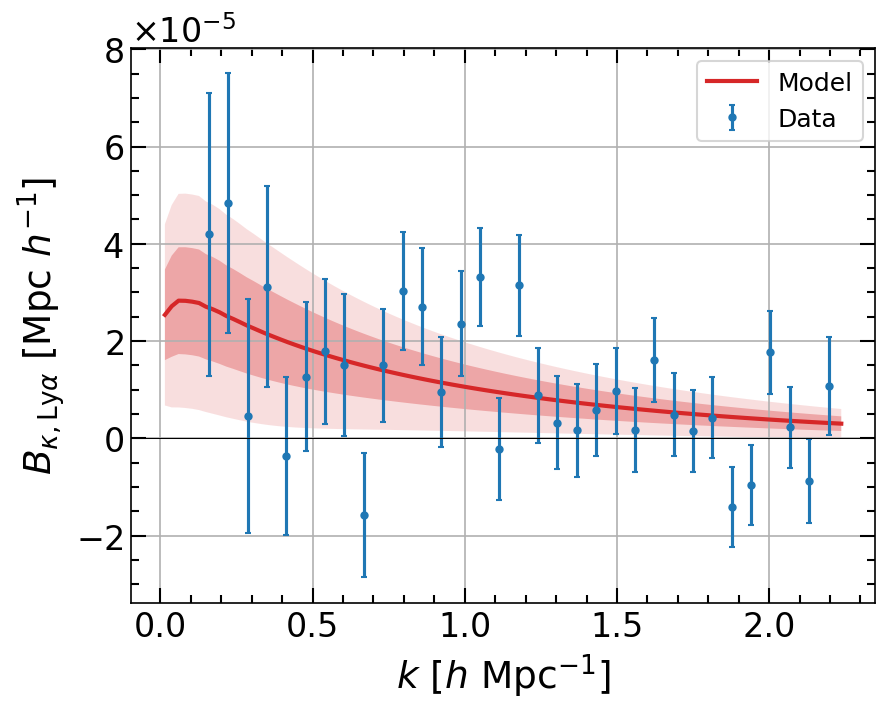}
    \caption{\label{fig:bestfit}Best-fit vs data, where $\chi^2 / \nu = 42.0 / (33-3)$, which is a reasonable value. However, some data points are near outliers by visual inspection. These points could indicate missing nuisance parameters in our model or underestimated errors in our bootstrap covariance matrix.}
\end{figure}

\begin{figure}
    \centering
    \includegraphics[width=\columnwidth]{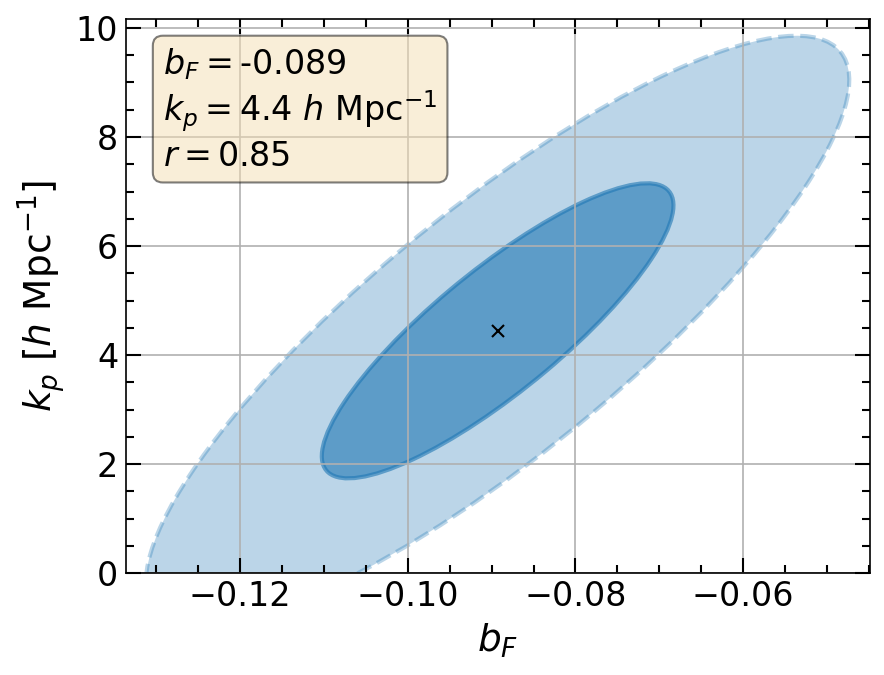}
    \caption{\label{fig:bf_kp_contour}One and two sigma contours for $b_F-k_p$ based on the minimizer covariance matrix. We find $b_F=-0.09 \pm 0.02$ and $k_p=4.4 \pm 2.7~$\hmpc at an effective redshift $z_\mathrm{eff}=2.44$, and the correlation coefficient between them is $r=0.85$.}
\end{figure}

Figure~\ref{fig:bestfit} compares the best-fit model to the data points.
We find $b_F=-0.09 \pm 0.02$, $k_p=4.4 \pm 2.7~$\hmpc\ and $\log T = 4.0 \pm 2.1$ at an effective redshift $z_\mathrm{eff}=2.44$. Both $b_F$ and $k_p$ are highly correlated with correlation coefficient $r=0.85$, which can be seen in Figure~\ref{fig:bf_kp_contour}, whereas $k_p$ and $\log T$ are less correlated with $r=0.1$. The $\log T$ constraints are not better than our priors, so even though these correlations are expected, the numerical values we report are affected by the priors.
\citet{bourbouxCompletedSDSSIVExtended2020} reports a \lya\ velocity bias of $b_{\eta}=-0.2014$ at $z_\mathrm{eff}=2.334$, which can be converted to the flux bias $b_F = b_\eta f / \beta = -0.116 \pm 0.005$, where $f=0.97$ is the growth rate. So our best-fit $b_F$ value is in the right vicinity, but $1.5\sigma$ away from the literature value. Furthermore, these best-fit parameters yield $\chi^2 / \nu = 42.0 / (33-3)$, which is a reasonable value.

\subsection{Position-dependent framework}
\citet{douxFirstDetectionCosmic2016} provides an intuitive theoretical model for \bispeconed\ based on the location-dependent power spectrum and the response of the power spectrum to a large-scale density mode. Their model has a quadratic dependence on the Wiener filter, whereas our model has a linear dependence. \citet{douxFirstDetectionCosmic2016} erroneously define the large-scale density mode based on the Wiener filter. However, this filter is a post-processing step based on the instrumental noise properties of the CMB experiment, so the power spectrum cannot respond to it (unless the quasar spectra are angularly smoothed with the same Wiener filter). Following \citet{chiangLyaCmbXmodel2018}, the correct formulation for the location-dependent modeling is as follows:
\begin{align}
    \label{eq:response}P_\mathrm{1D}(k, \textbf{x}) &= P^\mathrm{lin}_\mathrm{1D}(k) + \overline{\delta}(\textbf{x}) \partderiv{P_\mathrm{1D}}{\overline{\delta}} \\
    \langle P_\mathrm{1D}(k, \textbf{x}) \kappa(\textbf{x}) \rangle &= \langle \overline{\delta}(\textbf{x}) \kappa(\textbf{x}) \rangle \partderiv{P_\mathrm{1D}}{\overline{\delta}},
\end{align}
where the ``survey window function" $W_L$ for each quasar is a line in space, so the large-scale density mode is a simple average in the radial direction without any angular smoothing component:  $\overline{\delta}(\textbf{x}) = \frac{1}{\Delta \chi}\int d\chi\ \delta_m(\chi, \bm{x}) W_L(\chi - \chi_*)$, where $W_L$ is one within the \lya\ forest region and zero otherwise \citep{chiangPositiondependentPowerSpectrum2014}. Then, the variance term is given by
\begin{align}
    \langle \overline{\delta}(\textbf{x}) \kappa(\textbf{x}) \rangle &= \int \frac{d\chi}{\Delta \chi} \frac{d^3 \bm{p}}{(2\pi)^3} e^{-i p_z (\chi - \chi_*)} \nonumber \\
    &\qquad \times W_\kappa(\chi)  W_\mathrm{WF}(p_\bot \chi) \Tilde{W}_L(p_z) P(p) \\
    \langle \overline{\delta}(\textbf{x}) \kappa(\textbf{x}) \rangle &\approx W_\kappa(\chi_*) \int \frac{d^2 \bm{p}_\bot}{(2\pi)^2} W_\mathrm{WF}(p_\bot \chi_*) P(p_\bot),
\end{align}
where we apply the Limber approximation in the last line and recover the linear dependence on the Wiener filter. The response term $\partial P_\mathrm{1D} / \partial \overline{\delta}$ can be calculated with the standard perturbation theory approach as described in \citet{douxFirstDetectionCosmic2016}.
\begin{equation}
    \partderiv{P_\mathrm{1D}}{\overline{\delta}}(k_z) = \int \frac{d^2 \bm{k}_\bot}{(2\pi)^2} b_F^2 \left(1 + \beta_F \mu^2\right)^2 D(k, \mu) f(k) P_L(k),
\end{equation}
where $k = \sqrt{k_\bot^2 + k_z^2}$, $\mu = k_z / k$, $D(k, \mu)$ is the non-linear fitting function for the \lya\ forest \citep{mcdonaldPredictingLyaPower2003, arinyoNonLinearPowerLya2015} and $f(k)$ is the response of the linear matter power spectrum \citep{chiangPositiondependentPowerSpectrum2014}:
\begin{equation}
    f(k) = \frac{68}{21} - \frac{1}{3} \frac{d \ln k^3 P_L(k)}{d \ln k}.
\end{equation}

As noted in \citet{douxFirstDetectionCosmic2016}, this formulation of $\partial P_\mathrm{1D} / \partial \overline{\delta}$ ignores the response of bias and other parameters to the large-scale density mode. Alternatively, the response can be numerically derived from separate universe simulations \citep{chiangResponseToSqueezedBisQsoLyaX2017, chiangLyaCmbXmodel2018}. Furthermore, as noted by \citet{chiangResponseToSqueezedBisQsoLyaX2017}, Eq.~\ref{eq:response} will have additional terms related to velocity bias and tidal field.

\section{\label{sec:discuss}Discussion}
\subsection{Measurement concerns}
\bispeconed\ faces the same challenges as \poned, originating with the quasar continuum fitting algorithm, HCDs, metals, and pipeline noise and resolution estimates. We partially addressed the errors in the quasar continuum fitting, HCDs, and metal contamination in Section~\ref{subsec:results}.  Fortunately, \bispeconed\ is unbiased against noise systematics as instrumental noise and $\kappa$ are uncorrelated. However, scatter in noise errors between quasars will inflate the covariance of \bispeconed. We applied a CCD amplifier-dependent correction to mitigate this effect as described in Section~\ref{subsec:contfit}. The spectrograph resolution errors will introduce a scale-dependent bias to \bispeconed\ since the resolution correction is multiplicative. This effect is relevant at higher $k$ values than we measure, so we ignore it in this work.

Additionally, our continuum fitting method biases the quasar continuum estimates towards the average transmission in the forest, which introduces a small bias that affects all scales. For \poned\ measurements, this bias is at most 1\% at low redshifts ($z < 2.6$). Since our measurements are not as precise, we ignore the scaling of \poned\ in \bispeconed. However, the correlations between these biased continuum errors and lensing convergence could be more important for \bispeconed. \citet{chiangLyaCmbXmodel2018} calculated that these correlations could constitute 30\% of the total signal at $z=2.2$ and become comparable to the true signal at $z=2.6$. Due to spectral noise, the fitted continuum amplitude is not as correlated with the underlying large-scale density field as assumed in \citet{chiangLyaCmbXmodel2018}, so these should be considered upper bounds.

\begin{figure}
    \centering
    \includegraphics[width=\columnwidth]{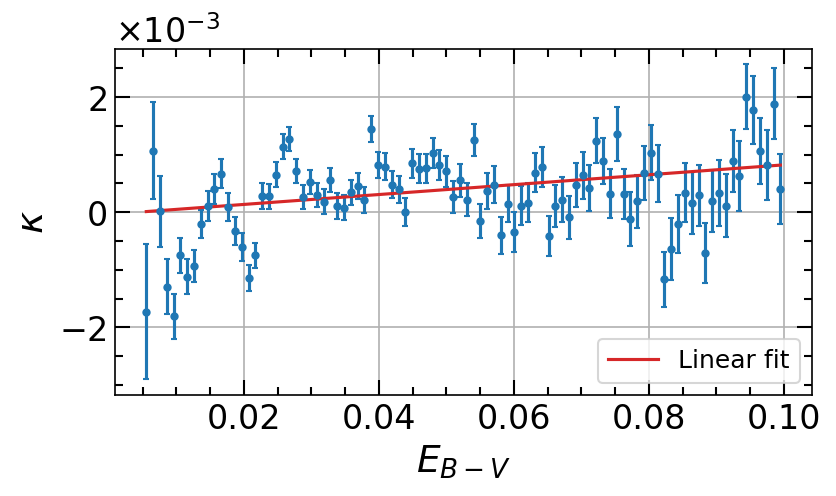}
    \caption{\label{fig:ebv_kappa_rel}$\kappa-E_{B-V}$ relation. The best linear fit is $\alpha=0.01$. As an additional test, we also compare DESI $E_{B-V}$ values to the corrected SFD with CIB removed dust map $E_{B-V}$ values from \citet{chiangCorrectedSFD2023}, which yields a smaller bias value of $\alpha = 0.005$. These weak biases do not affect our measurement.}
\end{figure}

Another possible correlated contamination between DESI quasar spectra and the Planck gravitational lensing map is the Galactic extinction. This is because Galactic extinction corrections are applied to both the data used for quasar target selection and the CMB measurements. We calculate the $B_{E, \mathrm{Ly}\alpha}$ bispectrum between \poned\ and $E_{B-V}$ by replacing $\kappa$ with $E_{B-V}$ values in the DESI DR1 quasar catalog \citep[SFD, ][]{schlegelMapOfDustInfrared1998}. We then quantify the correlation between $\kappa$ and $E_{B-V}$ by binning quasar $\kappa$ values with respect to $E_{B-V}$. Figure~\ref{fig:ebv_kappa_rel} shows our measured $\kappa-E_{B-V}$ relation.
For a linear bias relation of $\kappa = \alpha E_{B-V} + \mathrm{signal}$, \bispeconed\ will have a dust contamination term proportional to $\alpha$ such that $\Delta$\bispeconed$=\alpha B_{E, \mathrm{Ly}\alpha}$. We do not detect $B_{E, \mathrm{Ly}\alpha}$ at any significance, but find that it is numerically comparable to \bispeconed. However, we also find that $\alpha=0.01$ and this weak correlation between $\kappa$ and the Galactic extinction does not change our bispectrum measurement and does not increase our error budget. As an additional test, we compare DESI $E_{B-V}$ values to the corrected SFD (cSFD) with Cosmic Infrared Background (CIB) removed dust map values from \citet{chiangCorrectedSFD2023} and find  $\delta E_{B-V} \sim 0.001$. cSFD dust map has a better correction for the imprint of the large-scale structure and yields a smaller bias value between the gravitational lensing and Galactic extinction ($\alpha = 0.005$). For stronger correlations or more precise future bispectrum measurements, the $\kappa-E_{B-V}$ bias model can be generalized to any polynomial such that:
\begin{align}
    \Delta \kappa &= \sum_{n=1} \alpha_n E_{B-V}^n \\
    \Delta B_{\kappa, \mathrm{Ly}\alpha} &= \sum_{n=1} \alpha_n B^{(n)}_{E, \mathrm{Ly}\alpha},
\end{align}
where $B^{(n)}_{E, \mathrm{Ly}\alpha}$ is the bispectrum between \poned\ and $E^n_{B-V}$. Alternatively, this $\kappa-E_{B-V}$ relation can be used to de-bias $\kappa$ values. The angular correlations between Galactic extinction and gravitational lensing can produce spurious correlations. They may require a more careful treatment when measuring the angular $\delta_F-\kappa$ cross-power spectrum and cross-bispectrum.

To test for contamination from clusters in the foreground, we also measure \bispeconed\ using the thermal Sunyaev-Zeldovich (tSZ) effect-deprojected lensing maps. These are temperature-only (TT) estimates.  We find both \bispeconed\ completely agree within the error bars with slight downward trend at low $k$, but the detection significance becomes $4.1\sigma$ due to lost signal.

We also test the Planck PR4 lensing map to gauge the improvements in the new pipeline \citep{carronCmbLensingPr4Maps2022}. We find a small increase ($\approx 0.2\sigma$) in detection significance, which is not significant enough to repeat the analysis with this new reduction. The Atacama Cosmology Telescope (ACT) map has an even lower reconstruction noise power but covers a smaller area than Planck \citep{madhavacherilActDr6CmbLensing}, so the net gain will be similarly incremental.

\subsection{Modeling concerns}
Modeling \lya\ \poned\ is a challenging task since it requires sufficient accuracy at small scales to model the nonlinearities involved in gravitational collapse and hydrodynamics, yet also substantial volumes to capture the full extent of modern surveys. The generation of hydrodynamical simulations that incorporate all these effects is computationally expensive. One promising approach is to use emulators that are trained on existing simulations to capture the essential mapping between cosmological parameters and the corresponding \poned \citep{pedersenEmulator2021, fernandezMultifidelityEmulator2022,  cabayolNeuralNetworkLyaEmu2023}. This approach essentially trades the free bias parameters $b_F, \beta_F$ with external observations of the IGM mean flux $\overline{F}$, and gains additional constraining power. Similar emulators for \bispeconed\ will be powerful in advancing cosmological interpretation.

With increased precision in \bispeconed, metal contributions such as Si~\textsc{iii} and Si~\textsc{ii} oscillations may need to be included in the model.

A further modeling complication is the He~\textsc{ii} reionization at $z \approx 3$ \citep{gaikwadConsistentRobustMeasurement2020}, and the residual ionization bubbles. The beginning, strength, and duration of the reionization will be related to the overall density in a given region. The net effect of this inhomogeneous process will likely add power to $k>0.01$\skm\ modes \citep{upton_sanderbeckInhomogeneousHeII2020}.

\section{\label{sec:summary}Summary}
We presented a 1D CMB lensing--\lya\ forest bispectrum measurement using DESI DR1 quasar spectra and the Planck PR3 lensing map. This cross-bispectrum measurement is robust against some instrumental systematics and has a $\sigma_8^4$ dependence, compared to the $\sigma_8^2$ dependence of the power spectrum, which could help break the $b_F-\sigma_8$ degeneracy. This is the second measurement of its kind and our $4.8\sigma$ detection is comparable to the first detection presented by \citet{douxFirstDetectionCosmic2016} at $5\sigma$. We found that the calibration of the pipeline noise estimates based on CCD amplifier regions reduced the errors in \bispeconed\ by reducing the scatter in noise systematics from quasar to quasar. This calibration produced an improvement in our detection significance of about $0.7\sigma$.

We performed a shuffle test to confirm the signal was real and then isolated the signal to the neutral hydrogen in the intergalactic medium by subtracting the metal contamination. We measured $\kappa-$metals bispectrum using the sideband technique and found that the metal contamination was not insignificant. With this additional uncertainty, our detection significance decreased to $2.7\sigma$.

We investigated how Galactic extinction and clusters of galaxies could be foreground contaminants that correlate with the  \lya\ forest from DESI and $\kappa$ from Planck. We found that these foregrounds are not strong enough to affect \bispeconed\ at our current precision. The angular correlations within these foregrounds could complicate future angular $\kappa-$\lya\ cross-power spectrum and cross-bispectrum measurements.

We developed a theoretical model within the Limber and flat-sky approximations and calculated it using tree-level perturbation theory. Our model provided a reasonable fit, but it lacks the second-order bias terms such as $b_2$ and nuisance parameters for additional non-linear enhancements and angular dependent effects. The effective field theory of the \lya\ forest is an encouraging avenue for future study \citep{ivanovEffectiveLya2024}.

In the future, \bispeconed\ can be used to constrain bias parameters and break the degeneracy between $\sigma_8$ and other cosmological parameters. This requires sizable improvements in the data quality of both the \lya\ forest and the reconstructed CMB lensing maps.

\section*{Data availability}
All data points shown in the figures are available in simple text files on the following website: \url{https://doi.org/10.5281/zenodo.11152068}. DESI DR1 will be publicly available in the future.

\begin{acknowledgments}
NGK thanks Christopher Hirata for the helpful discussion on numerical integration of the tree-level perturbation theory.

NGK and PM acknowledge support from the United States Department of Energy, Office of High Energy Physics under Award Number DE-SC-0011726.

This material is based upon work supported by the U.S. Department of Energy (DOE), Office of Science, Office of High-Energy Physics, under Contract No. DE–AC02–05CH11231, and by the National Energy Research Scientific Computing Center, a DOE Office of Science User Facility under the same contract. Additional support for DESI was provided by the U.S. National Science Foundation (NSF), Division of Astronomical Sciences under Contract No. AST-0950945 to the NSF’s National Optical-Infrared Astronomy Research Laboratory; the Science and Technology Facilities Council of the United Kingdom; the Gordon and Betty Moore Foundation; the Heising-Simons Foundation; the French Alternative Energies and Atomic Energy Commission (CEA); the National Council of Humanities, Science and Technology of Mexico (CONACYT); the Ministry of Science and Innovation of Spain (MICINN), and by the DESI Member Institutions: \url{https://www.desi.lbl.gov/collaborating-institutions}.

The DESI Legacy Imaging Surveys consist of three individual and complementary projects: the Dark Energy Camera Legacy Survey (DECaLS), the Beijing-Arizona Sky Survey (BASS), and the Mayall z-band Legacy Survey (MzLS). DECaLS, BASS and MzLS together include data obtained, respectively, at the Blanco telescope, Cerro Tololo Inter-American Observatory, NSF’s NOIRLab; the Bok telescope, Steward Observatory, University of Arizona; and the Mayall telescope, Kitt Peak National Observatory, NOIRLab. NOIRLab is operated by the Association of Universities for Research in Astronomy (AURA) under a cooperative agreement with the National Science Foundation. Pipeline processing and analyses of the data were supported by NOIRLab and the Lawrence Berkeley National Laboratory. Legacy Surveys also uses data products from the Near-Earth Object Wide-field Infrared Survey Explorer (NEOWISE), a project of the Jet Propulsion Laboratory/California Institute of Technology, funded by the National Aeronautics and Space Administration. Legacy Surveys was supported by: the Director, Office of Science, Office of High Energy Physics of the U.S. Department of Energy; the National Energy Research Scientific Computing Center, a DOE Office of Science User Facility; the U.S. National Science Foundation, Division of Astronomical Sciences; the National Astronomical Observatories of China, the Chinese Academy of Sciences and the Chinese National Natural Science Foundation. LBNL is managed by the Regents of the University of California under contract to the U.S. Department of Energy. The complete acknowledgments can be found at \url{https://www.legacysurvey.org/}.

Any opinions, findings, and conclusions or recommendations expressed in this material are those of the author(s) and do not necessarily reflect the views of the U. S. National Science Foundation, the U. S. Department of Energy, or any of the listed funding agencies.

The authors are honored to be permitted to conduct scientific research on Iolkam Du’ag (Kitt Peak), a mountain with particular significance to the Tohono O’odham Nation.
\end{acknowledgments}

\appendix
\section{\label{app:qmle}Quadratic estimator}
The QMLE works in real space (instead of Fourier space) to estimate the power spectrum, and can weight pixels by the pipeline noise and the intrinsic \lya\ large-scale structure correlations. In this section, we provide a summary of our previous work \cite{karacayliOptimal1DLy2020, karacayliOptimal1dDesiEdr2023}.

Our QMLE implementation adopts a fiducial power spectrum $P_{\mathrm{fid}}(k, z)$ to calculate the signal contribution of 1D \lya\ forest correlations of the covariance matrix.
\begin{equation}
    \label{eq:pd13_fitting_fn}\frac{kP_{\mathrm{fid}}}{\pi} = A \frac{(k/k_0)^{3 +n + \alpha\ln k/k_0}}{1+(k/k_1)^2} \left(\frac{1+z}{1+z_{0}}\right)^{B + \beta\ln k/k_0},
\end{equation}
where $k_{0} = 0.009~$\skm\ and $z_{0}=3.0$. We stress that this functional form is sufficient to weight pixels, but does not capture all of the scientific information in \poned. The parameters we used in this analysis are the best-fit values to the DESI early data \poned\ measurement \cite{karacayliOptimal1dDesiEdr2023} and are listed in Table~\ref{tab:app:pfidparams}.

\begin{table}
\caption{\label{tab:app:pfidparams}
Fiducial parameters
}
\begin{ruledtabular}
\begin{tabular}{lccc}
 & $A$ & $n$ & $\alpha$ \\
\colrule
\lya\ & 0.076 & -2.52 & -0.13 \\
SB~1 & 0.0021 & -3.07 & -0.074\\
\colrule\\
& $B$ & $\beta$ & $k_1$ [\skm] \\ 
\colrule
\lya\ & 3.67 & 0.29 & 0.037 \\
SB~1 & 1.60 & -0.24 & 0\\
\end{tabular}
\end{ruledtabular}
\end{table}

Given a collection of pixels representing normalized flux fluctuations $\bm{\delta}_F$, the quadratic estimator is formulated as follows:
\begin{equation}
    \label{eq:theta_it_est}\hat \theta_{a} = \sum_{b} \frac{1}{2} F^{-1}_{ab}(d_{b} - n_{b} - t_{b}),
\end{equation}
where
\begin{align}
    \label{eq:data_dn} d_a &= \bm{\delta}_F^\mathrm{T} \mathbf{C}^{-1}\mathbf{Q}_a \mathbf{C}^{-1} \bm{\delta}_F, \\
    \label{eq:noise_bn}n_a &= \Tr(\mathbf{C}^{-1}\mathbf{Q}_a \mathbf{C}^{-1}\mathbf{N}), \\
    \label{eq:signalfid_tn}t_a &= \Tr(\mathbf{C}^{-1}\mathbf{Q}_a \mathbf{C}^{-1}\mathbf{S}_{\mathrm{fid}}).
\end{align}
The covariance matrix $\mathbf C \equiv \langle\bm{\delta}_F\bm{\delta}_F^T\rangle$ is the sum of signal and noise, $\mathbf C = \mathbf{S}_{\mathrm{fid}} + \mathbf{N}$. Note that this is the covariance matrix between forest pixels at different wavelengths. The quantity $\mathbf{Q}_a \equiv \partial \mathbf{C} / \partial \theta_a$ and the estimate of the Fisher matrix is
\begin{equation}
    \label{eq:fisher_matrix}F_{ab} = \frac{1}{2} \Tr(\mathbf{C}^{-1}\mathbf{Q}_a \mathbf{C}^{-1} \mathbf{Q}_b ).
\end{equation}
Assuming different quasar spectra are uncorrelated, the Fisher matrix $F_{ab}$ and the expression in parentheses in Eq.~\ref{eq:theta_it_est} can be computed for each quasar, then accumulated, i.e. $\mathbf{F}=\sum_q\mathbf{F}_{q}$, etc. The power spectrum covariance matrix noted in the text is the inverse of this Fisher matrix, i.e. $\mathbf{C}_q = \mathbf{F}^{-1}_q$.

\section{\label{app:derive_N}Derivation of sample bias correction}
The bispectrum estimator in Eq.~\ref{eq:bispec_estimator} is biased without the sample correction term $\mathbf{N}$. We derive this bias by calculating the expected value of the biased estimator $\hat{B}^*_a$.
\begin{align}
    \hat{B}^*_a =& \sum_q W^q_{ab} \left(p^q_{b} - \hat{\overline{{p}}}_b \right) \left(\kappa^q - \hat{\overline{{\kappa}}}_a \right) \\
    =& \sum_q W^q_{ab} p^q_{b} \kappa^q - \sum_q \kappa_q W^q_{ab} \hat{\overline{{p}}}_b \nonumber\\
    &- \sum_q  W^q_{ab} p^q_{b} \hat{\overline{{\kappa}}}_a  + \sum_q  W^q_{ab} \hat{\overline{{p}}}_b \hat{\overline{{\kappa}}}_a \label{eq:app:bis}
\end{align}
We use the Einstein summation convention such that all repeated indices imply summation \textbf{except} for the index $a$.
The third and fourth terms cancel regardless of the choice for the mean $\kappa$ estimation method.
Then:
\begin{align}
    \hat{B}^*_a =& \sum_q W^q_{ab} p^q_{b} \kappa^q - \sum_q W^q_{ab} \hat{\overline{{p}}}_b \kappa^q \\
    =& \sum_q W^q_{ab} p^q_{b} \kappa_q - \sum_{q, q'} W^q_{ab} W^{q'}_{bc} p^{q'}_{c} \kappa^q \\
    =& \sum_q W^q_{ab} p^q_{b} \kappa^q - \sum_{q} W^q_{ab} W^q_{bc} p^q_{c} \kappa^q \nonumber\\
    &- \sum_{q, q'\neq q} W^q_{ab} W^{q'}_{bc} p^{q'}_{c} \kappa^q \\
    \langle \hat{B}^*_a \rangle =& \sum_q W^q_{ab} \langle p \kappa\rangle_b - \sum_{q} W^q_{ab} W^q_{bc} \langle p \kappa\rangle_c \nonumber\\
    &- \sum_{q} W^q_{ab} \sum_{q'\neq q}W^{q'}_{bc} \langle{p}\rangle_{c} \langle{\kappa}\rangle,
\end{align}

\begin{figure}
    \centering
    \includegraphics[width=\columnwidth]{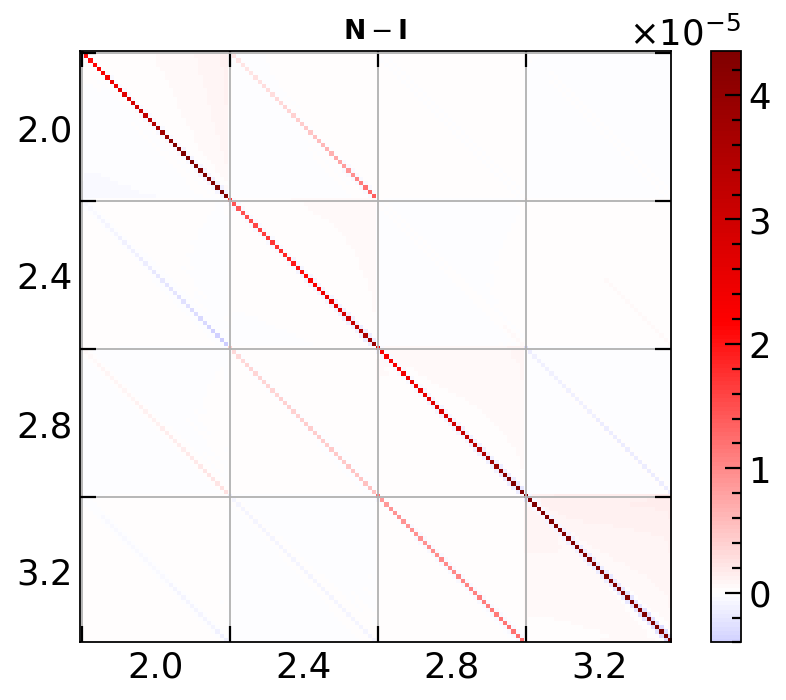}
    \caption{\label{fig:samplebias}Identity subtracted sampling bias matrix $N_{ab} - \delta^K_{ab}$. This correction is small as expected from a large sample.}
\end{figure}

Using the relation $\sum_{q'\neq q} \mathbf{W}_{q'} = \mathbf{I} - \mathbf{W}_{q}$, we can collect all the terms into a simple expression
\begin{equation}
    \langle \hat{B}^*_a \rangle = \left(\mathbf{I} - \sum_q \mathbf{W}^2_{q}\right)_{ab} \left(\langle p \kappa\rangle_b - \langle{p}\rangle_{b} \langle{\kappa}\rangle \right),
\end{equation}
and read $\mathbf{N}^{-1} = \mathbf{I} - \sum_q \mathbf{W}^2_{q}$. This sampling bias correction term is small as expected from a large sample, which is shown in Figure~\ref{fig:samplebias}.
Note that this matrix is not symmetric.

\section{\label{app:theory_derivation}Derivation of bispectrum}
We can derive the bispectrum expression without assuming a constant lensing kernel in a redshift bin:
\begin{align}
    \kappa^\mathrm{WF}(\chi, \bm{x}) =& \int \mathrm{d}\chi \ W_\kappa(\chi) \delta^\mathrm{WF}_m(\chi, \bm{x}), \\
    \delta^\mathrm{WF}_m(\chi, \bm{x}) =& \int \frac{d^2\bm{p}_\bot}{(2\pi)^2} e^{i\bm{p}_\bot \cdot \bm{x}} W_\mathrm{WF}(p_\bot \chi) \Tilde{\delta}_m(\chi, \bm{p}_\bot),
\end{align}
where we still applied the flat-sky approximation. The 3D Fourier transform of $\delta^\mathrm{WF}_m$ is modified accordingly:
\begin{equation}
    \delta^\mathrm{WF}_m(\chi, \bm{x}) = \int \frac{d^2\bm{p}_\bot dp_z}{(2\pi)^3} e^{i (\bm{p}_\bot \cdot \bm{x} + p_z \chi)} W_\mathrm{WF}(p_\bot \chi) \Tilde{\delta}_m(\bm{p}).
\end{equation}

The bispectrum calculation then proceeds as follows:
\begin{align}
    &\langle \delta_F(k, \bm{x}) \delta_F(k', \bm{x}) \kappa(\bm{x}) \rangle\nonumber \\
    & \quad = \int \mathrm{d}\chi\ W_\kappa(\chi) \langle \delta_F(k, \bm{x}) \delta_F(k', \bm{x}) \delta^\mathrm{WF}_m(\chi, \bm{x}) \rangle \\
    & \quad = \int \mathrm{d}\chi \int \frac{d^2 \bm{q}_\bot d^2\bm{q}'_\bot d^2\bm{p}_\bot}{(2\pi)^6} e^{i (\bm{q}_\bot + \bm{q}'_\bot + \bm{p}'_\bot)\cdot \bm {x} }\int  \frac{dp_z}{2\pi} e^{i p_z \chi} \nonumber \\
    &\quad\qquad \times W_\kappa(\chi) W_\mathrm{WF}(p_\bot \chi) B^\mathrm{3D}_{FFm}(\bm{q}, \bm{q}', \bm{p}) \nonumber \\
    &\quad\qquad \times (2\pi)^2 \delta_D(\bm{q}_\bot + \bm{q}'_\bot + \bm{p}_\bot) \nonumber \\
    &\quad\qquad \times (2\pi) \delta_D(k + k' + p_z).
\end{align}
We can integrate out $\bm{q}'_\bot$ as before. The integration over $p_z$ now yields $p_z = -k - k'$ due to the Dirac delta function. Then:
\begin{align}
    &\langle \delta_F(k, \bm{x}) \delta_F(k', \bm{x}) \kappa(\bm{x}) \rangle\nonumber \\
    & \quad = \int \frac{d^2 \bm{q}_\bot d^2 \bm{p}_\bot}{(2\pi)^4} B^\mathrm{3D}_{FFm}(\bm{q}, \bm{q}', \bm{p}) \nonumber \\
    & \quad \quad \times \int \mathrm{d}\chi\ e^{-i (k + k') \chi} W_\kappa(\chi) W_\mathrm{WF}(p_\bot \chi)
\end{align}

So far we only used the flat-sky approximation. Now we apply the Limber approximation by assuming the integrand slowly varies in $\chi$:
\begin{align}
    & \int \mathrm{d}\chi\ e^{-i (k + k') \chi} W_\kappa(\chi) W_\mathrm{WF}(p_\bot \chi) \nonumber \\
    & \qquad \approx (2\pi) \delta_D(k+k') \left[W_\kappa(\chi_*) W_\mathrm{WF}(p_\bot \chi_*) + O_1 \right]
\end{align}
and recover the expression in the main text. A first-order correction to the Limber approximation is given by
\begin{align}
    O_1 = \frac{(\Delta z)^2}{8}\chi''_* [& W'_\kappa(\chi_*) W_\mathrm{WF}(p_\bot \chi_*) \nonumber \\
    & + p_\bot W_\kappa(\chi_*) W'_\mathrm{WF}(p_\bot \chi_*)],
\end{align}
which yields 0.1\% corrections on \bispeconed\ for $\Delta z =0.4$ at $z_*=2.4$.

To include the angular modes in our model, we start from the following definition:
\begin{equation}
    \langle \delta_F(k, \bm{x}) \delta_F(k', \bm{x}) \kappa(\bm{x}') \rangle = 2\pi \delta_D(k+k') \zeta(k, r),
\end{equation}
where $r = |\bm{x}' - \bm{x}|=\theta\chi_*$ and the bispectrum is given by the Fourier transform:
\begin{equation}
    B^\mathrm{2D}_{FFm}(k, p_\bot) = \int d^2\bm{r} e^{i \bm{p}_\bot \cdot \bm{r}} \zeta(k, r).
\end{equation}
Under the same flat-sky and Limber approximations, we can then obtain the following:
\begin{align}
    &B^\mathrm{2D}_{FFm}(k, p_\bot) = W_\kappa(\chi_*) W_\mathrm{WF}(p_\bot \chi_*) \nonumber \\
    & \qquad \times \int \frac{q_\bot dq_\bot d\phi}{(2\pi)^2} B^\mathrm{3D}_{FFm}(\bm{q}, \bm{q}', \bm{p}),
\end{align}
where $\bm{q}_\bot \cdot \bm{p}_\bot = q_\bot p_\bot \cos\phi$ in $B^\mathrm{3D}_{FFm}$ calculation. Note that $B^\mathrm{1D}_{FFm}(k)$ is equivalent to $\zeta(k, r=0)$. By applying the inverse Fourier transform, we can recover the previous $B^\mathrm{1D}_{FFm}$ expressions.
\begin{align}
    \zeta(k, r) &= \int \frac{d^2\bm{p}_\bot}{(2\pi)^2} e ^{-i \bm{p}_\bot\cdot \bm{r}} B^\mathrm{2D}_{FFm}(k, p_\bot) \\
    B^\mathrm{1D}_{FFm}(k) &= \int \frac{d^2\bm{p}_\bot}{(2\pi)^2} B^\mathrm{2D}_{FFm}(k, p_\bot) \\
    &=  \int \frac{p_\bot dp_\bot}{2\pi} B^\mathrm{2D}_{FFm}(k, p_\bot)
\end{align}

\bibliography{references}

\end{document}